\documentstyle[aps,multicol,epsf]{revtex}

\begin{document}
\draft

\title{Dynamics of ripple formation in sputter erosion: nonlinear 
phenomena \\}
\author{S. Park$^1$, B. Kahng$^{1,2}$, H. Jeong$^2$, and 
A.-L. Barab\'asi$^2$\\} 
\address{$^1$ Department of Physics and Center for Advanced 
Materials and Devices, Konkuk University, Seoul 143-701, Korea\\
$^2$ Department of Physics, University of Notre Dame, 
Notre Dame, IN 46556\\}
\maketitle 
\begin{center}
%(\today)
\end{center}

\thispagestyle{empty}

\begin{abstract} 
Many morphological features of sputter
eroded surfaces are determined by the balance between ion induced 
linear instability and surface diffusion. 
However, the impact of the nonlinear terms on the 
morphology is less understood.  
We demonstrate that while at short times ripple formation 
is described by the linear theory, after a characteristic time 
the nonlinear terms determine the surface morphology
by either destroying the ripples, or generating a new rotated ripple 
structure. We show that the morphological 
transitions induced by the nonlinear effects can 
be detected by monitoring the surface 
width and the erosion velocity. 
\end{abstract} 
\begin{multicols}{2}
\narrowtext
The morphological evolution of ion sputtered surfaces has generated much 
experimental and theoretical interest in recent years.
As a result, there is extensive evidence that  
ion bombardment can result in ordered surface ripples, 
or lead to kinetic roughening, depending on the experimental conditions. 
These experimental results, which cover amorphous and crystalline 
materials (SiO$_2$\cite{chason1}), and both metals (Ag\cite{exp}) and 
semiconductors (Ge\cite{chason2}, Si\cite{carter,aziz}), have motivated 
extensive theoretical work aiming to uncover the mechanism 
responsible for ripple formation and kinetic roughening. 
A particularly successful model has been proposed by 
Bradley and Harper (BH) \cite{harper},
in which the height $h(x,y,t)$ of the eroded surface is 
described by the linear equation 
\begin{equation}
{\partial_t h} = \nu_x \partial_x^2 h +\nu_y \partial_y^2 h 
-K\partial^4 h,
\end{equation}
where $K$ is the surface diffusion constant and the coefficients 
$\nu_x$ and $\nu_y$ are induced by the 
erosion process such that $\nu_y < 0$ and $\nu_x$ can change 
sign as the angle of the incidence of the ions is varied.
The balance of the unstable erosion term ($-|\nu| \partial^2 h$) 
and the smoothening surface diffusion term ($-K\partial^4 h$) 
generates ripples with wavelength
\begin{equation}
\ell_i=2\pi \sqrt{2K/|\nu_i|}, 
\end{equation}
where $i$ refers to the direction ($x$ or $y$) along 
which the associated $\nu_i$ ($\nu_x$ or $\nu_y$) is 
the largest. While successful in predicting 
the ripple wavelength and orientation\cite{koponen},  
this linear theory cannot explain a number 
of experimental features, such as the saturation 
of the ripple amplitude\cite{wittmaack,aziz10,vajo}, 
the observation of rotated ripples\cite{rot_exp}, 
and the appearance of kinetic roughening\cite{eklund,yang}. 
Recently it has been proposed \cite{cuerno} that the 
inclusion of nonlinear terms 
and noise (both of which were derived from Sigmund's theory of 
sputtering\cite{sigmund}) 
can cure these shortcomings. Consequently, 
Eq.(1) has to be replaced by the noisy Kuramoto-Sivashinsky 
equation (KS) equation\cite{kuramoto} 
\begin{eqnarray}
{\partial_t h} &=& \nu_x \partial_x^2 h +\nu_y \partial_y^2 h 
-K_x\partial_x^4 h -K_y \partial_y^4 h-K_{xy}\partial_x^2 
\partial_y^2 h \nonumber \\
& &+{\lambda_x \over 2}(\partial_x h)^2 + {\lambda_y \over 2} 
(\partial_y h)^2 + \eta(x,y,t), 
\end{eqnarray}
where $\eta(x,y,t)$ is an uncorrelated white noise with zero mean, 
mimicking the randomness resulting from the stochastic nature of 
ion arrival to the surface. 
All coefficients in Eq.(3) have been determined in terms of the experimental 
parameters, such as the ion flux, angle of incidence, ion penetration 
depth, and substrate temperature\cite{cuerno,maxim1}.\\ 

While it is expected that the nonlinear theory incorporates 
most features of ripple formation and kinetic roughening, the morphological 
and dynamical features of the surface described by it are known 
only in certain special cases. 
For example, when the nonlinear terms and the noise are neglected 
($\lambda_x=\lambda_y=0$, $\eta=0$), Eq.(3) reduces to the 
linear theory (1), and predicts ripple formation. 
It is also known that the isotropic KS equation 
($\nu_x=\nu_y < 0$, $K_x=K_y=K_{xy}/2$, and  $\lambda_x=\lambda_y$) 
asymptotically (for large time and length scales) predicts kinetic 
roughening, with exponents similar to that seen experimentally 
in ion sputtering\cite{eklund}.
For positive $\nu_x$ and $\nu_y$, Eq.(3) reduces to the anisotropic 
KPZ equation\cite{kpz}, whose scaling behavior is controlled 
by the sign of $\lambda_x \cdot \lambda_y$\cite{wolf}. 
Finally, recent integration by Rost and Krug \cite{rost} 
of the noiseless version of Eq.(3) provided evidence 
that when $\lambda_x \cdot \lambda_y < 0$, new ripples, 
unaccounted for by the linear theory, appear and their direction is 
rotated with respect to the ion direction\cite{rost}. 
However, it is not known if this rotated phase survives 
in the presence of noise, $\eta$. 
These special cases illustrate the complexity 
of the morphological evolution predicted by Eq.(3). 
To be able to make specific predictions on the morphology of ion-sputteted 
surfaces, we need to gain a full understanding of the behavior predicted 
by (3), going beyond the special cases, that are often experimentally 
irrelevant. In this paper, we integrate numerically Eq.(3), 
aiming to uncover the dynamics and the morphology of the surfaces 
for different values of the parameters. 
We demonstrate a clear separation of the linear and nonlinear 
behavior. For short erosion times, the ripple development 
and orientation follow the predictions of the linear 
theory of BH. However, after a well defined crossover time, 
which depends on the coefficients of Eq.(3), the surface 
morphology is determined by the nonlinear terms. We find 
that when $\lambda_x \cdot \lambda_y > 0$ 
the nonlinear terms destroy the ripple morphology. However, 
when $\lambda_x \cdot \lambda_y < 0$, they result in 
a long and apparently rough transient regime, followed by 
a new morphology of rotated ripples, as seen in the noiseless 
KS equation. We show that these morphological transitions 
can be detected by monitoring the surface width or erosion 
velocity, quantities that can be measured more easily 
$in$ $situ$. Finally, we discuss the impact of our result 
on current and future experimental work.\\ 

The direct numerical integration is carried out by discretizing 
the continuum equation of (3), using the standard discretization 
techniques\cite{numerical}. 
We choose a temporal increment 
$\Delta t=0.01$ and impose periodic boundary conditions 
$h(x,y,t)=h(x+L,y,t)=h(x,y+L,t)$ where $L \times L$ 
is the size of the substrate.  
We choose the noise to be uniformly distributed between [-1/2,1/2], 
and perturb the initial flat configuration with the noise.
Since the sign of the nonlinear terms plays 
a significant role in defining the surface morphology,
we discuss separately the $\lambda_x \cdot \lambda_y > 0 $ and 
$\lambda_x \cdot \lambda_y < 0$ cases.\\
 
$\lambda_x \cdot \lambda_y > 0$ ---  
A general feature of sytems such as Eq.(3) is that the nonlinear terms  
do not affect the surface morphology or dynamics until 
a crossover time $\tau$ has been reached. 
Thus, we expect that for early times, i.e. for $t < \tau$, the 
surface morphology and dynamics is properly 
described by the linear theory. To demonstrate this separation 
of the linear and nonlinear regimes, in Fig.~1 
we show the time dependence of the surface width defined as   
$W^2 (L,t) \equiv {1 \over L^2}\sum_{x,y} h^2(x,y,t)-{\bar h}^2$ 
and of the mean height $\bar h ={1 \over L^2}\sum_{x,y} h(x,y,t)$. 
We find that for $t < \tau$, the width $W$ increases exponentially 
while the mean height stays constant at $\bar h=0$. 
Indeed, both of these findings are consistent with 
the predictions of the linear theory: 
$W$, being proportional to the ripple amplitude, according to Eq.(1) 
increases as $W \sim \exp(\nu t/\ell^2)$, and the linear terms 
do not change the mean height of the surface. 
Furthermore, inspecting the surface morphology, we 
find that in this regime the ripple wavelength and 
orientation are also correctly described by the linear theory.  
For example, 
for the parameters $\nu_x=-0.0001$, $\nu_y=-0.6169$, $K_x=K_y=K_{xy}/2=2$, 
and $\lambda_x=\lambda_y=-0.001$,
according to (2), the ripple wavelength along the $y$ axis 
is $\ell_y \approx 16$, and along the $x$-axis is $\ell_x \approx 1257$. 
Since the dominant wavelength is determined by the growth rate 
$\sim \exp(\nu t/\ell^2)$, the smaller wavelength, i.e. $\ell_x$, 
will dominate.
As Fig.~2a shows, for a system of size $64 \times 64$ we observe 
four ripples aligned along the $x$-axis, in agreement with the 
previous prediction.  
As a second example we consider the case,  
$\nu_x=-1.2337$, $\nu_y=-0.0001$, 
$K_x=K_y=K_{xy}/2=1$, and $\lambda_x=\lambda_y=-0.001$, 
for which we expect ripples of wavelength $\ell_x \approx 8$, 
smaller than $\ell_y \approx 889$.  
As Fig.~2b shows, in this case we observe eight ripples aligned along 
the $y$-axis. \\

While the early time behavior is correctly predicted by the linear theory, 
beyond the crossover time $\tau$ the nonlinear terms become effective. 
One of the most striking consequence of these terms is that the surface width 
stabilizes rather abruptly (see Fig.~1). Furthermore, 
the ripple pattern generated in the linear regime disappears, 
and the surface exhibits kinetic roughening. 
A typical surface morphology, demonstrating the absence of ripples, is 
shown in Fig.~2c. 
The crossover time $\tau$ from the linear to the nonlinear behavior 
can be estimated by comparing the strength of the linear term with that 
of the nonlinear term.  
Let the typical height at the crossover time $\tau$ 
be $W_0\equiv \sqrt{W^2(L,\tau)}$. 
Then, from the linear equation we obtain 
$W_0 \sim \exp(\nu \tau /\ell^2)$, 
while from $\partial_t h \sim \lambda (\partial h)^2$ we 
estimate $W_0/\tau \sim \lambda W_0^2/\ell^2$. 
Combining these two relations we obtain 
\begin{equation}
\tau \sim (K/\nu^2)\ln (\nu/\lambda). 
\end{equation}
In this expression, $\nu$, $K$ and $\lambda$, refer to 
the direction perpendicular to the ripple orientation. 
The predicted $\lambda$-dependence of $\tau$ is confirmed 
in the inset of Fig.~1a. 
An another quantity that refects the transition from the linear 
to the nonlinear regime is the erosion velocity 
$v=\partial_t {\bar h}$. The main contribution to 
the erosion velocity comes from a constant erosion 
rate $-v_0$, that has been omitted from (1) and 
(3), since it does not affect the surface morphology 
\cite{harper,cuerno}.  
However, in addition to $v_0$, the mean height is also 
modified by the nonlinear terms, $\lambda_x (\partial_x h)^2$ 
and $\lambda_y (\partial_y h)^2$.  
In the following for simplicity, we neglect the $v_0$ term, since 
its value does not depend on the surface morphology, 
and it is constant throughout the erosion process.
The nonlinear terms act to decrease the mean height 
in the case of $\lambda_x < 0$ and 
$\lambda_y < 0$. We can estimate the surface velocity as 
$v \sim \lambda W_0^2 /\ell^2\sim \nu^3/(K\lambda)$ 
using $W_0 \sim \nu/\lambda$. 
This dependence of $v$ on $\lambda$ is consistent with 
the numerical results, shown in the inset of Fig.~1b. 
In this regime ($t > \tau$) the surface exhibits kinetic 
roughening, i.e. the surface width should increase either 
logarithmically (when $\lambda=0$) or as a power law 
(when $\lambda \ne 0$)\cite{wolf,fractal}. 
However, compared with the exponential 
increase in the early regime ($t < \tau$), this dependence 
is hardly observable. The simulation times required to investigate 
the asymptotic scalings of $W$ with $t$ are currently 
prohibitive.\\

$\lambda_x \cdot \lambda_y < 0$ ---
As Fig.~3a shows, we again observe a 
separation of the linear and nonlinear regimes, however, 
we find that the morphology and the dynamics of the surface 
in the nonlinear regime is quite different from the 
case $\lambda_x \cdot \lambda_y > 0$. 
In regime I, for early times ($t < \tau$), 
the surface forms ripples (see Fig.~4a), 
whose wavelength and orientation is correctly described by 
the linear theory. After the first crossover 
time $\tau$, given by Eq.~(4), 
the surface width is stabilized, 
and the ripples disappear, as shown in Fig.~4b. 
After $\tau$, the system enters a rather long transient 
regime, that we call the regime II. 
Here, the surface is rough, and no 
apparent spatial order is present. We often observe the development 
of individual ripples, but they soon disappear, and no 
long-range order is present in the system.  
However, at a second crossover time $\tau_2$, 
a new ripple structure suddenly forms, as shown 
in Fig.~4c, in which the ripples are stable and rotated 
with an angle $\theta_c$ to the $x$ direction. 
The angle $\theta_c$ has the value 
$\theta_c =\tan^{-1}\sqrt{{-\lambda_x/\lambda_y}}$, 
(or $\tan^{-1}\sqrt{{-\lambda_y/\lambda_x}}$)\cite{rost}, 
which can be calculated by moving to a rotated frame of 
coordinates that vanishes the nonlinear term in the transverse direction. 
Indeed, as Fig.~4c shows, the observed angle is in excellent agreement with 
$\theta_c=\tan^{-1}(1/2)$ for $\lambda_x=1$ and 
$\lambda_y=-4$. 
We also find that the time the system spends in regime II 
fluctuates from system to system, thus $\tau_2$ has a wide distribution. 
The transitions between the three regimes can be detected 
by monitoring the surface width $W$ (Fig.~3a):
in regime I, the width increases exponentially, as predicted 
by the linear theory; it is approximately constant but hightly 
fluctuating in regime II, and suddenly increases and stabilizes in 
regime III. 
Note that the amplitude of the rotated ripples in regime 
III is much larger than in regime II, a rather 
attractive feature for possible applications as 
patterned templates for various microelectronic applications.\\
 
The demonstrated morphological transitions generate an anomalous 
behavior in $\bar h$ as well. 
As Fig.~3a shows, the mean height is zero in the linear regime, 
increases as the ripples are destroyed in regime II, and 
decreases with a constant velocity in regime III. 
In order to understand this complex behavior, we consider 
a specific example, 
for which the surface morphologies are shown in Fig.~4. 
For this parameter set, ripples are aligned along the $y$-axis 
in the region I, because $\ell_x \ll \ell_y$. 
Thus, the contribution of $(\partial_x h)^2$ 
is much larger than that of $(\partial_y h)^2$, 
even though $|\lambda_x| < |\lambda_y|$, and 
the surface height increases due to the term 
$\lambda_x (\partial_x h)^2$ with $\lambda_x >0$ in regime II.  
However, as the ripples are destroyed by the nonlinear 
effects, the contribution of $(\partial_y h)^2$ term increases, 
and eventually $\lambda_y (\partial_y h)^2$ becomes 
larger than $\lambda_x(\partial_x h)^2$, forcing 
the mean height to decrease because $\lambda_y < 0$. 
The velocity in regime III is determined by the nonlinear 
coefficient in the direction along the ripples, which 
reduces to $\lambda_x+\lambda_y$ 
after the coordinate transformation to the rotated ripple 
direction. This prediction is in good agreement with 
the results of Fig.~4, which demonstrates that 
$v \sim 1/(\lambda_x+\lambda_y)$. \\

The clear separation of the linear and the nonlinear behavior, 
that holds for both signs of $\lambda_x \cdot \lambda_y$, 
has a direct impact on the experimental observations. 
Numerous experiments have observed the development of ripples 
whose wavelength and orientation is in good agreement with 
the prediction of the linear 
theory \cite{chason1,exp,chason2,aziz,koponen}. 
Based on our results, we expect that these 
experiments were in the $t < \tau$ regime, 
where indeed the linear theory fully describes 
the system. However, recent results have provided 
detailed experimental evidence of ripple amplitude 
stabilization \cite{wittmaack,aziz10,vajo}, a clear sign 
of the presence of nonlinear effects. 
Furthermore, it was found that the different $W$ versus $t$ curves 
can be collapsed by rescaling time with a factor 
$\nu^2/K$ and amplitude with $\sqrt{\nu/2K}$\cite{aziz10}. 
Indeed, this is in excellent agreement with our prediction, Eq.(4). 
Finally, since the values of $\nu$ and $\lambda$ can be 
tuned by changing the ion energy and the angle 
of incidence, and $K$ can be tuned with the temperature, 
the values of $\tau$ and $\tau_2$ can be 
changed continuously, and thus our predictions on 
the morphological transition between the linear 
and nonlinear regimes could be tested experimentally.  
Furthermore, the detailed morphological evolution uncovered 
here, combined with earlier calculations that connect 
the coefficients in Eq.(3) to the numerical values 
of the parameters describing the ion-bombardment process 
\cite{harper,cuerno,maxim1}, offer a detailed roadway 
that can guide further experiments and 
facilitate the use of ion sputtering for surface 
patternings.\\ 

This work is supported by KOSEF (Grant No. 971-0207-025-2), 
the Korean Research Foundation (Grant No. 98-015-D00090), 
NSF-DMR and ONR.

\begin{figure}
\centerline{\epsfxsize=6.8cm \epsfbox{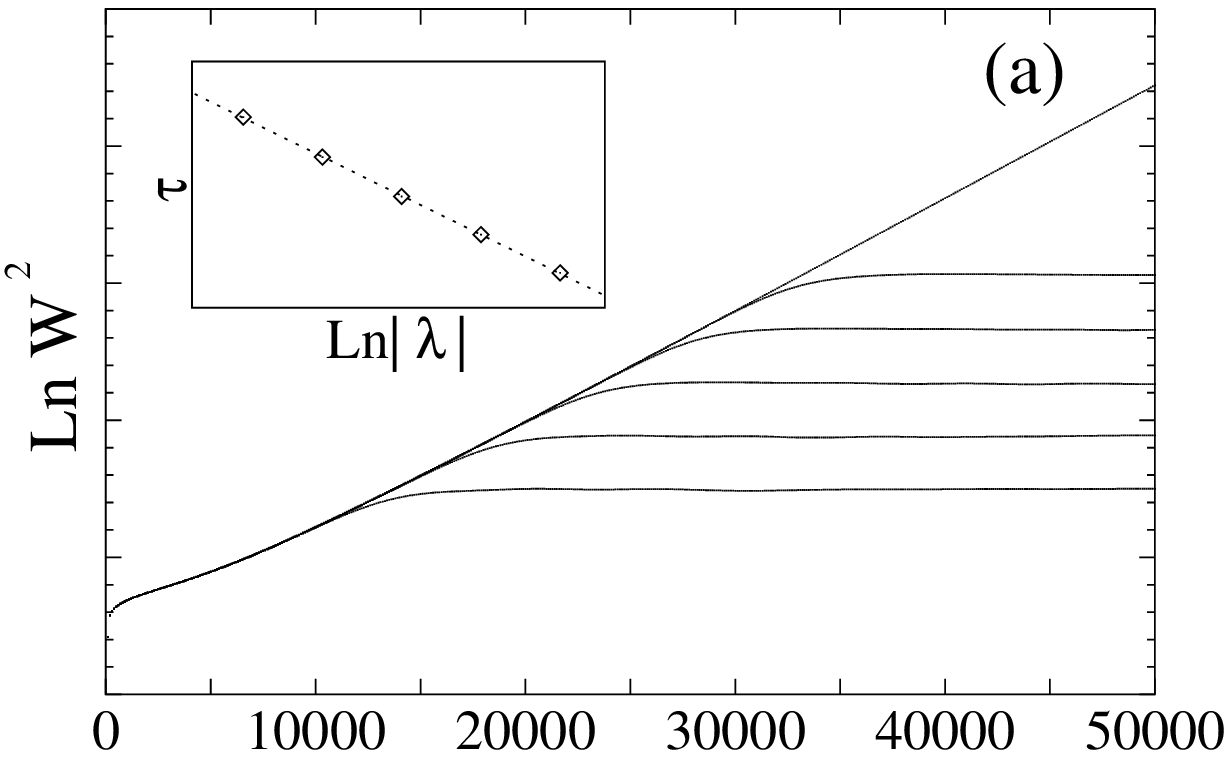}}
\centerline{\epsfxsize=6.8cm \epsfbox{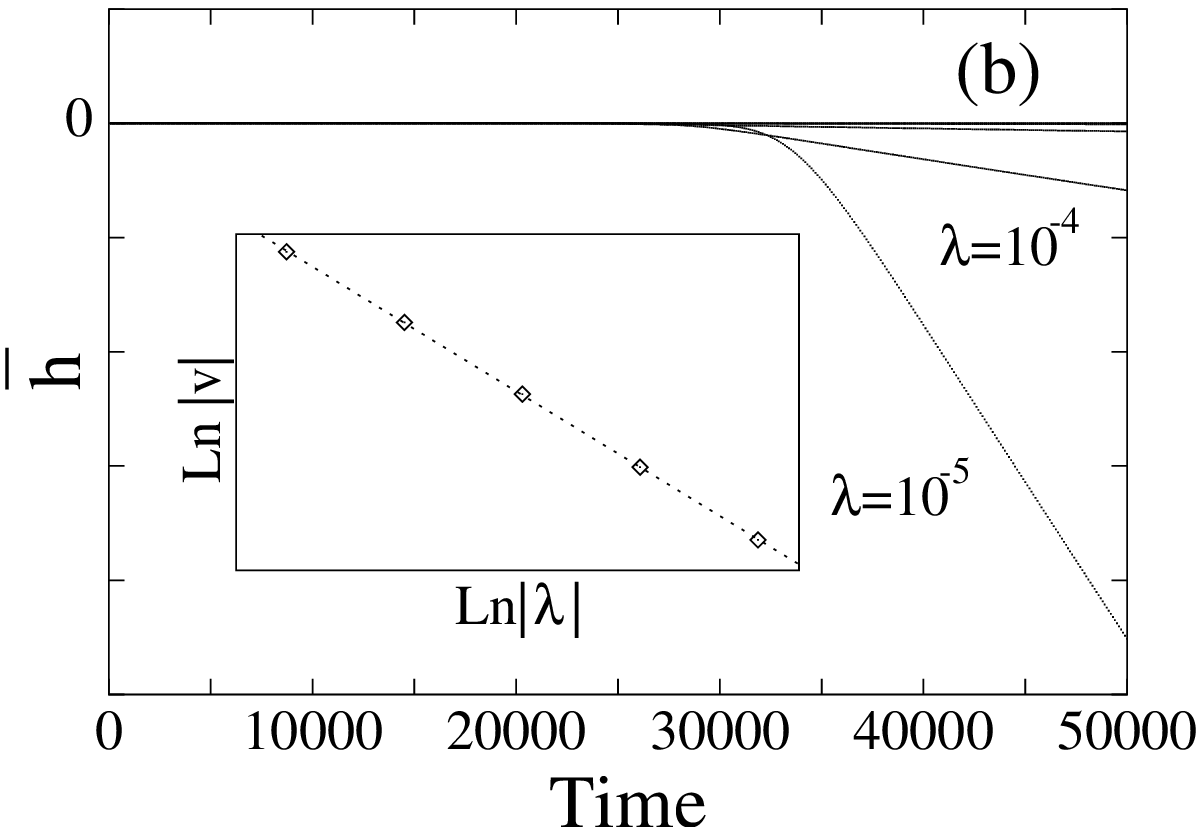}}
\caption{Time evolution of (a) the surface width $W^2$  
and (b) the mean height $\bar h$ for the 
parameters $\nu_x=-0.0001$, $\nu_y=-0.6169$, 
$K_x=K_y=K_{xy}/2=2$. The different curves correspond 
to different values of $\lambda_x=\lambda_y=\lambda$. 
In (a), from top to bottom, the curves correspond to 
$\lambda=0$, $-10^{-5}$, $-10^{-4}$, 
$-10^{-3}$, $-10^{-2}$, and $-10^{-1}$, respectively.
In (b), from bottom to top, they correspond to $\lambda=-10^{-5}$, 
$-10^{-4}$, $-10^{-3}$, $-10^{-2}$, and $-10^{-1}$, respectively.
Inset (a): The crossover time $\tau$, estimated from (a) 
is shown as a function of $\ln |\lambda|$. 
Inset (b): Plot of $\ln |v|$ versus $\ln |\lambda|$. 
The dotted line has a slope $\approx -1.07$, implying $v \sim 
1/\lambda$.}
\label{fig1}
\end{figure}

\begin{figure}
\centerline{\epsfxsize=8.3cm \epsfbox{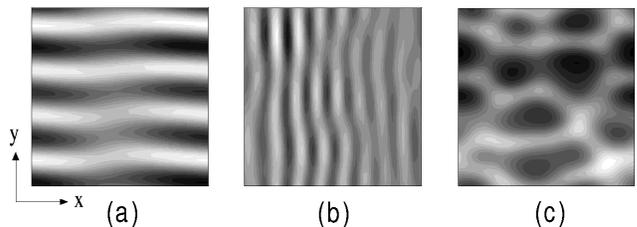}}
\caption{Grey scale plot of a surface of size $64 \times 64$ 
with the parameter sets: (a) $\nu_x=-0.0001$, $\nu_y=-0.6169$, 
$K_x=K_y=K_{xy}/2=2$, and $\lambda_x=\lambda_y=-0.001$ 
at $t=2 \times 10^4$. (b) $\nu_x=-1.2337$, 
$\nu_y=-0.0001$, $K_x=K_y=K_{xy}/2=1$ and 
$\lambda_x=\lambda_y=-0.001$ at $t=2 \times 10^3$. 
Eq.~(2) predicts the wavelength $\ell_y \approx 16$ for (a), 
and $\ell_x \approx 8$ for (b). 
(c) The surface configuration at time $t=5 \times 10^5$ 
for the parameter set used in (a).}
\label{fig2}
\end{figure}

\begin{figure}
\centerline{\epsfxsize=8.5cm \epsfbox{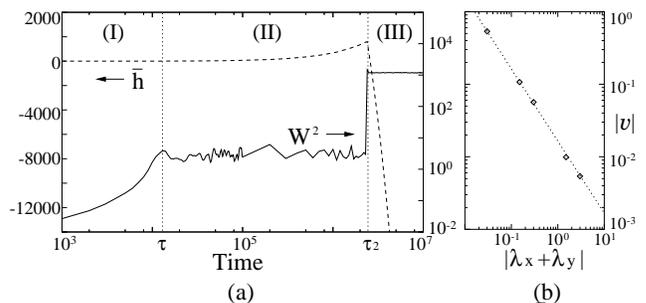}}
\caption{(a) Time evolution of the mean height $\bar h$ 
(dashed, left linear scale) and the surface 
width (solid, right logarithmic scale) 
for the parameters,  
$\nu_x=-0.6169$, $\nu_y=-0.01$, $K_x=K_y=K_{xy}/2=2$, 
$\lambda_x=1$ and $\lambda_y=-4$. 
The dotted lines seperate the three regimes discussed in the 
text.
(b) The dependence of $|v|$ on the nonlinear terms 
$|\lambda_x+\lambda_y|$ for the same parameters used in 
(a). The dotted line has a slope $\approx -1.02$, 
implying $v \sim 1/(\lambda_1 +\lambda_2)$.}   
\label{fig3}
\end{figure}

\begin{figure}
\centerline{\epsfxsize=8.3cm \epsfbox{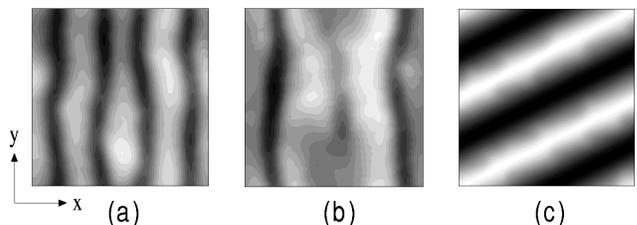}}
\caption{Grey scale plot of a surface 
of size $64\times 64$ with the parameter set used in Fig.~3, 
showing the surface morphologies at $t=10^4$ in 
regime I (a), $t=2 \times 10^5$ in regime II (b), and $t=10^7$ 
in regime III (c).}
\label{fig4}
\end{figure}

\end{multicols}
\end{document}